\documentclass[11pt]{article}
\usepackage{amsmath,amssymb}
\usepackage{amscd}
\def\hs{\hbox to 3mm{}}
\def\hhs{\hbox to 5cm{}}
\def\ss{\smallskip}

\def\si{\sigma}

\def\al{\alpha}
\def\ia{\rightarrow\hspace{-13pt}+\ }
\def\ca{\nearrow\hspace{-10pt}\searrow}
\def\cs{\rightarrow\hspace{-15pt}+\ }
\def\SGS{\bigoplus_{n\geq0}{\mathbb Q}[\SG_n]}
\def\FQSym{{\bf FQSym}}
\def\MQSym{{\bf MQSym}}
\def\Zy{\frak Z}
\def\F{\mathbf{F}}
\def\G{\mathbf{G}}

\def\ncs#1#2{#1\langle\langle #2 \rangle\rangle}

\def\SG{\frak{S}}

\def\N{\mathbb{N}}

\def\Z{\mathbb{Z}}
\def\p{\frak p}

\newtheorem{example}{Example}[section]
\newtheorem{note}[example]{Note}
\newtheorem{theorem}[example]{Theorem}

\newtheorem{proposition}[example]{Proposition}

\def\mref#1{(\ref{#1})}
\def\Proof{\medskip\noindent {\it Proof --- \ }}
\def\cqfd{\hfill$\Box$ \bigskip}

\title{\scshape{ Free quasi-symmetric functions, product actions and quantum field theory of partitions.}}
\author{
{\bf G\'erard H. E. Duchamp $^\dag$, Jean-Gabriel Luque $^\ddag$, K A Penson $^{\star}$
Christophe Tollu$^\dag$},\\ ($^\dag$) LIPN, UMR CNRS 7030\\
Institut Galil\'ee - Universit\'e Paris-Nord\\ 99, avenue
Jean-Baptiste Cl\'ement\\ 93430 Villetaneuse, France\\ ($^\ddag$)
IGM,   Laboratoire d'informatique UMR 8049 IGM-LabInfo\\ 77454
Marne-la-Vall\'ee Cedex 2, France.\\ $^{\star}$ LPTL, 
Laboratoire   de  Physique   Th\'{e}orique  des  Liquides, CNRS UMR 7600\\
Universit\'{e} Pierre et Marie Curie,\\
Tour 16, $5^{i\grave{e}me}$ \'{e}tage, 4, place Jussieu, F 75252 Paris Cedex 05,
France\\
{\tt
\{ghed@lipn.univ-paris13.fr, luque@univ-mlv.fr,}\\ 
{\tt penson@lptl.jussieu.fr, ct@lipn.univ-paris13.fr\}}}

\begin{document}
\maketitle 
\begin{quote}{\footnotesize {\bf Abstract}: We examine two associative
products over the ring of symmetric functions related to the
intransitive and Cartesian products of permutation groups. As an
application, we give an enumeration of some Feynman type diagrams
arising in Bender's QFT of partitions.
We end by exploring possibilities to construct noncommutative analogues.}\end{quote}

\begin{quote}{\footnotesize {\bf R\'esum\'e}: Nous \'etudions deux lois produits
associatives sur les fonctions sym\'etriques correspondant aux
produits intransitif et cartésien des groupes de permutations.
Nous donnons comme application l'\'enum\'eration de certains
diagrammes de Feynman apparaissant dans la QFT des
partitions de Bender. Enfin, nous
donnons quelques pistes possibles pour construire des analogues
non-commutatifs.}\end{quote}
\section{Introduction}

In a relatively recent paper, Bender, Brody and Meister introduce
a special Field Theory described by
\begin{equation}\label{QFT0}
G(z)=\Big(e^{(\sum_{n\geq 1}L_n \frac{z^n}{n!}\frac{\partial}{\partial x})}\Big) \Big(e^{(\sum_{m\geq 1}V_m \frac{x^m}{m!})}\Big)
\Big|_{x=0}
\end{equation}
in order to prove that any sequence of numbers $\{a_n\}$
can be generated by a suitable set of rules applied to
some type of Feynman diagrams \cite{BBM1,BBM2}. These diagrams actually are bipartite finite graphs
with no isolated vertex, and edges
weighted with integers.\\

Expanding one factor of \mref{QFT0}, we can observe surprising
links between: the normal ordering problem (for bosons), the parametric
Stieltjes moment problem and the convolution of kernels, substitution
matrices (such as generalised Stirling matrices) and one-parameter
groups of analytic substitutions \cite{Benin, JPA, SymS}.

\ss
The aim of this paper is to make explicit the multifaceted connections
between noncommutative symmetric functions (here \MQSym, \FQSym\,
\cite{nc6}) and the Feynman diagrams arising in the expansion
of formula \mref{QFT0} used in combinatorial physics \cite{SymS}.

\ss
The structure of the contribution is the following. In Section
\ref{S2}, we define two associative products in
$\SG=\bigsqcup\SG_n$ related to the Intransitive and Cartesian products
of permutation groups. These products induce a structure of
$2$-associative algebra over the symmetric functions. The
properties of this algebra are investigated in Section \ref{S3}. At the
end of this section, we give, as an application, an inductive
formula for computing generating series of Bender's Feynman
diagrams. Noncommutative analogues are proposed in Section
\ref{S4}.

\section{Actions of a direct product of permutation
groups}\label{S2}
\subsection{Direct product actions\label{DP}}
The actions of the direct product of two permutation groups  (in
particular,  the structure of the cycles) give rise to interesting
properties related to the enumeration of unlabelled objets \cite{Po}.
We open this section with the definition of two actions (namely,
Intransitive and Cartesian). For greater detail about these
constructions (or for constructions involving the wreath product)
the reader can refer to \cite{Cam1}.

Consider two pairs $(G_1,X_1)$ and $(G_2,X_2)$, where each $G_i$ is
a permutation group acting on $X_i$. The {\it intransitive action}
of $G_1\times G_2$ on $X_1\sqcup
 X_2$ (here $\sqcup$
means disjoint union) is defined by the rule
\begin{equation}
(\sigma_1,\sigma_2)x=\left\{\begin{array}{ll}\sigma_1x&\mbox{ if
}x\in X_1\\ \sigma_2x&\mbox{ if }x\in X_2
\end{array}\right. .
\end{equation}
This action  will be denoted by $(G_1,X_1)\ia(G_2,X_2):=(G_1\times
G_2,X_1\sqcup
 X_2)$.\\
 The {\it Cartesian action} of
$G_1\times G_2$ on $X_1\times X_2$ is defined by
\begin{equation} (\sigma_1,\sigma_2)(x_1,x_2)=(\sigma_1
x_1,\sigma_2 x_2).
\end{equation}
This action will be denoted by $(G_1,X_1)\ca(G_2,X_2):=(G_1\times
G_2,X_1\times X_2)$. Note that neither of the two laws just defined is
commutative. A natural question to ask is whether such a structure
enjoys some algebraic properties. For example, is the $\ca$ law
distributive over $\ia$?\\ In other words, what is the meaning of
\[(G_1,X_1)\ca((G_2,X_2)\ia(G_3,X_3))=(G_1\times G_2\times G_3,X_1\times(X_2\sqcup
 X_3))\]
 and
 \[((G_1,X_1)\ca(G_2,X_2))\ia((G_1,X_1)\ca(G_3,X_3))=(G_1\times G_2\times G_1\times G_3,(X_1\times
 X_2)\sqcup
(X_1\times X_3)).\]
 The groups $G_1\times G_2\times G_1\times G_3$ and $G_1\times G_2\times
 G_3$ are not isomorphic, so
 distributivity does not hold, although the set-theoretical Cartesian product is distributive over disjoint union.
However an examination of the
 structure of the cycles (see \cite{Cam1} for the general construction
 or section \ref{real} for a particular case) shows that the
 cycles are the same.
More precisely, a cycle can appear with different multiplicities
according to which group is acting, but
 if we focus on the set of the cycles, the two structures are
 similar.

 Now, let us give a construction which
takes such a phenomenon into account.
\subsection{Explicit realization}\label{real}

We will denote by
$\circ_N$ the natural action of $\SG_n$ on $\{0,\dots,n-1\}$.
Let $\SG_n$ and $\SG_m$ be two symmetric groups, we note  by
$\circ_I$ the intransitive action of $\SG_n\times \SG_m$ on
$\{0,\cdots,n+m-1\}$ and by $\circ_C$ the {\it Cartesian action} of
$\SG_n\times \SG_m$ on $\{0,\dots,nm-1\}$. More precisely,
\begin{equation}
(\sigma_1,\sigma_2)\circ_Ii=\left\{\begin{array}{ll}
\sigma_1\circ_Ni&\mbox{if }0\leq i\leq n-1\\
\sigma_2\circ_N(i-n)+n&\mbox{if }n\leq i\leq
n+m-1\end{array}\right. .
\end{equation}
for $0\leq i\leq n+m-1$, and

\begin{equation}
(\sigma_1,\sigma_2)\circ_C(j+nk)=(\sigma_1\circ_Nj)+n(\sigma_2\circ_Nk)
\end{equation}
for $0\leq j\leq n-1$ and $0\leq k\leq m-1$.

The {\it intransitive product} is the map
$\ia:\SG_n\times\SG_m\rightarrow \SG_{n+m}$ defined by
\begin{equation}
\sigma_1\ia\sigma_2=\sigma_1\sigma_2[n]
\end{equation}
where $\sigma_2[n]$ denotes $\sigma_2$ composed with the shifted substitution $i\rightarrow
i+n$ (here permutations are considered as words and $\ia$ is
nothing else but shifted concatenation).
\begin{example}\label{ex1} \rm Let $\sigma_1=1320\in\SG_4$ and
$\sigma_2=534120\in\SG_6$. Here, we denote a permutation of
$\SG_n$ by the word whose $i$th letter is the image of $i$ under
the natural action on $\{0,\dots,n-1\}$). With this notation, we
obtain
\begin{equation}
\sigma_1\ia\sigma_2=1320978564\nonumber
\end{equation}
and
\begin{equation}
\sigma_2\ia\sigma_1=5341207986\nonumber
\end{equation}
Clearly, it turns out that $\ia$ is not commutative.
\end{example}
The following proposition shows that the natural action of
$\SG_{n+m}$ coincides with the intransitive action of
$\SG_n\times\SG_m$.

\begin{proposition}\label{compint}
$(\sigma_1\ia\sigma_2)\circ_Ni=(\sigma_1,\sigma_2)\circ_Ii .$
\cqfd
\end{proposition}
Let us introduce a similar construction for the Cartesian action:
we define a map \\ $\ca:\SG_n\times\SG_m\rightarrow \SG_{nm}$
by
\begin{equation}\label{defcar}
\sigma_1\ca\sigma_2=\prod_{i,j}c_i\ca c'_j
\end{equation}
where $\sigma_1=c_1\cdots c_{k}$ (resp. $\sigma_2=c'_1\cdots c'_{k'}$)
is the decomposition of $\sigma_1$ (resp. $\sigma_2$) into a product
of cycles and \begin{equation}\label{cycleintr} c\ca
c'=\prod_{s=0}^{l\wedge l'-1}(\phi(s,0),\phi(s+1,1)\cdots,
\phi(s+l\vee l'-1,l\vee l'-1)),
\end{equation}
where $\wedge$ denotes the gcd,  $\vee$ denotes the lcm,
$c=(i_0,\cdots, i_{l-1})$, $c'=(j_0,\cdots,j_{l'-1})$ are two
cycles and $\phi(k,k')=i_{k\mbox{ mod }l}+nj_{k'\mbox{ mod }l'}$.
 Just like the Intransitive action, the Cartesian action coincides with the natural action.
\begin{proposition}\label{compcart}
$(\sigma_1\ca\sigma_2)\circ_Ni=(\sigma_1,\sigma_2)\circ_Ci\ .$
\end{proposition}
\Proof  From (\ref{defcar}), it suffices to prove the property
only when $\sigma_1=c$ and $\sigma_2=c'$ are two cycles. But as
(\ref{cycleintr}) is equivalent to \begin{eqnarray} c\ca
c'=&\displaystyle\prod_{s=0}^{l\wedge
l'-1}(i_s+nj_0,(c,c')\circ_C(i_s+nj_0),\dots,(c^{l\vee
l'-1},{c'}^{l\vee l'-1})\circ_C(i_s+nj_0))\nonumber\\
&=\displaystyle\prod_{s=0}^{l\wedge
l'-1}(i_s+nj_0,c\circ_Ci_s+nc'\circ_Nj_0 ,\dots,c^{l\vee
l'-1}\circ_Ni_s+n{c'}^{l\vee l'-1}\circ_Nj_0) ),\nonumber
\end{eqnarray} which completes the proof.\cqfd
\begin{example}\label{ex2} \rm
Consider the two permutations $\sigma_1=2031\in\SG_4$ and
$\sigma_2=01723456\in\SG_8$. The permutation $\sigma_1$ consists
of a unique cycle $c_1=(0,2,3,1)$ and $\sigma_2=c'_1c'_2c'_3$ is
the product of the three cycles $c'_1=(0)$, $c'_2=(1)$ and
$c'_3=(7,6,5,4,3,2)$. Hence, the permutation $\sigma_1\ca\sigma_2$
is the product of four cycles given by
\begin{enumerate}
\item $c_1\ca c'_1=(0,2,3,1)$
\item $c_1\ca c'_2=(4,6,7,5)$
\item $c_1\ca
c'_3=(28,26,23,17,12,10,31,25,20,18,15,9)(30,27,21,16,14,11,29,24,22,19,13,8).$
\end{enumerate}
To illustrate proposition \ref{compcart}, it suffices to draw the
cycles in the Cartesian product
$\{0,\dots,n-1\}\times\{0,\dots,m-1\}$ whose elements are re-
labelled $(i,j)\rightarrow i+nj$. For example, the two cycles
appearing in $c_1\ca c'_3$ give the following partition of
$\{0,1,2,3\}\times\{2,3,4,5,6,7\}$.
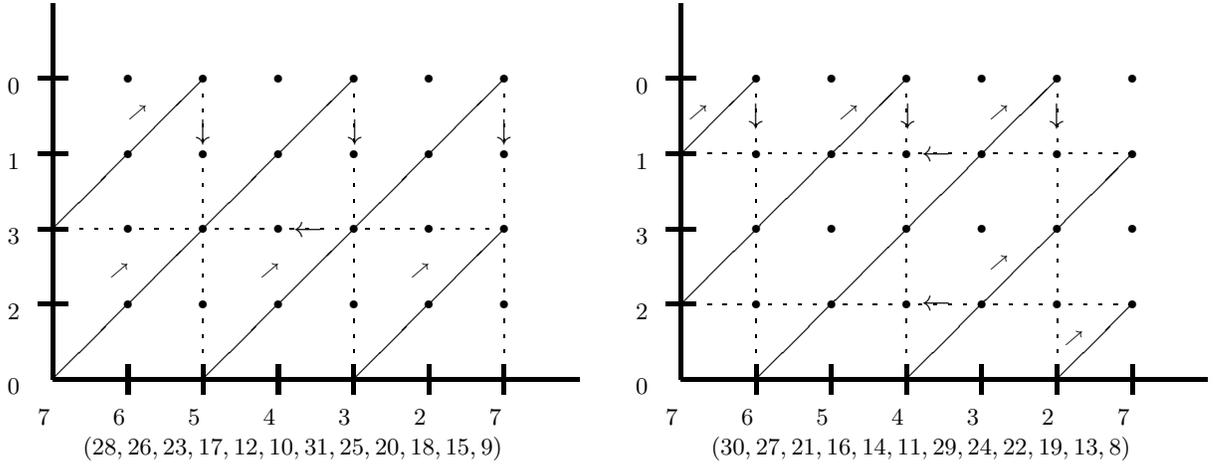
\begin{center}
\setlength{\unitlength}{2mm}
\begin{eqnarray}
\begin{picture}(45,30)(-5,-5)
\linethickness{1.5pt}
\put(0,0){\line(0,1){25}} \put(0,0){\line(1,0){35}}
\put(-1,5){\line(1,0){2}}\put(-1,10){\line(1,0){2}}
\put(-1,15){\line(1,0){2}}\put(-1,20){\line(1,0){2}}
\put(5,-1){\line(0,1){2}}\put(10,-1){\line(0,1){2}}
\put(15,-1){\line(0,1){2}}\put(20,-1){\line(0,1){2}}
\put(25,-1){\line(0,1){2}}\put(30,-1){\line(0,1){2}}
\put(-3,-1){\footnotesize 0}\put(-3,4){\footnotesize
2}\put(-3,9){\footnotesize 3} \put(-3,14){\footnotesize
1}\put(-3,19){\footnotesize 0}
\put(-1,-3){\footnotesize 7}\put(4,-3){\footnotesize
6}\put(9,-3){\footnotesize 5}\put(14,-3){\footnotesize
4}\put(19,-3){\footnotesize 3}\put(24,-3){\footnotesize 2}
\put(29,-3){\footnotesize 7}
\linethickness{1pt}
\put(0,0){\line(1,1){5}}\put(5,5){\line(1,1){5}}\put(10,10){\line(1,1){5}}
\put(15,15){\line(1,1){5}}\put(20,0){\line(1,1){5}}\put(25,5){\line(1,1){5}}
\put(0,10){\line(1,1){5}}\put(5,15){\line(1,1){5}}\put(10,0){\line(1,1){5}}
\put(15,5){\line(1,1){5}}\put(20,10){\line(1,1){5}}\put(25,15){\line(1,1){5}}
\linethickness{0.25pt} \multiput(20,0)(0,1){20}{\line(0,-1){0.2}}
\multiput(10,0)(0,1){20}{\line(0,-1){0.2}}
\multiput(30,0)(0,1){20}{\line(0,-1){0.2}}
 \multiput(30,10)(-1,0){30}{\line(-1,0){0.2}}
\put(5,17.5){{\tiny$\nearrow$}} \put(3.8,7){{\tiny$\nearrow$}}
\put(13.8,7){{\tiny$\nearrow$}} \put(23.8,7){{\tiny$\nearrow$}}
 \put(9.5,16){{$\downarrow$}}\put(19.6,16){{$\downarrow$}}
 \put(29.5,16){{$\downarrow$}}
  \put(16,9.5){{$\leftarrow$}}
  \put(2,-5){\footnotesize $(28,26,23,17,12,10,31,25,20,18,15,9)$}
 \multiput(5,5)(5,0){6}{\circle*{0.5}}
 \multiput(5,10)(5,0){6}{\circle*{0.5}}
 \multiput(5,15)(5,0){6}{\circle*{0.5}}
 \multiput(5,20)(5,0){6}{\circle*{0.5}}
\end{picture}&
\begin{picture}(45,30)(0,-5)
\linethickness{1.5pt}
\put(0,0){\line(0,1){25}} \put(0,0){\line(1,0){35}}
\put(-1,5){\line(1,0){2}}\put(-1,10){\line(1,0){2}}
\put(-1,15){\line(1,0){2}}\put(-1,20){\line(1,0){2}}
\put(5,-1){\line(0,1){2}}\put(10,-1){\line(0,1){2}}
\put(15,-1){\line(0,1){2}}\put(20,-1){\line(0,1){2}}
\put(25,-1){\line(0,1){2}}\put(30,-1){\line(0,1){2}}
\put(-3,-1){\footnotesize 0}\put(-3,4){\footnotesize
2}\put(-3,9){\footnotesize 3} \put(-3,14){\footnotesize
1}\put(-3,19){\footnotesize 0}
\put(-1,-3){\footnotesize 7}\put(4,-3){\footnotesize
6}\put(9,-3){\footnotesize 5}\put(14,-3){\footnotesize
4}\put(19,-3){\footnotesize 3}\put(24,-3){\footnotesize 2}
\put(29,-3){\footnotesize 7}
\linethickness{1pt}
\put(0,5){\line(1,1){15}}\put(15,0){\line(1,1){15}}
\put(0,15){\line(1,1){5}}\put(5,0){\line(1,1){20}}
\put(25,0){\line(1,1){5}}
\linethickness{0.25pt} \multiput(5,0)(0,1){20}{\line(0,-1){0.2}}
\multiput(15,0)(0,1){20}{\line(0,-1){0.2}}
\multiput(25,0)(0,1){20}{\line(0,-1){0.2}}
 \multiput(30,5)(-1,0){30}{\line(-1,0){0.2}}
  \multiput(30,15)(-1,0){30}{\line(-1,0){0.2}}
  \put(0.5,17.5){{\tiny$\nearrow$}}\put(10.5,17.5){{\tiny$\nearrow$}}
  \put(20.5,17.5){{\tiny$\nearrow$}} \put(20.5,7.5){{\tiny$\nearrow$}}
  \put(25.5,2.5){{\tiny$\nearrow$}}
  \put(4.5,17){{$\downarrow$}}\put(14.6,17){{$\downarrow$}}
  \put(24.5,17){{$\downarrow$}}
  \put(16,14.5){{$\leftarrow$}} \put(16,4.6){{$\leftarrow$}}
  \put(2,-5){\footnotesize $(30,27,21,16,14,11,29,24,22,19,13,8)$}
 \multiput(5,5)(5,0){6}{\circle*{0.5}}
 \multiput(5,10)(5,0){6}{\circle*{0.5}}
 \multiput(5,15)(5,0){6}{\circle*{0.5}}
 \multiput(5,20)(5,0){6}{\circle*{0.5}}
\end{picture}\nonumber
\end{eqnarray}
\end{center}
On the other hand,   the permutation $\sigma_2\ca\sigma_1$ is the
product of the four cycles
\begin{enumerate}
\item $c'_1\ca c_1=(0,16,24,8)$
\item $c'_2\ca c_1=(1,17,25,9)$
\item $c'_3\ca
c_1=(7,22,29,12,3,18,31,14,5,20,27,10)(6,21,28,11,2,23,30,13,4,19,26,15)$
\end{enumerate}
Clearly, $\sigma_1\ca\sigma_2\neq\sigma_2\ca\sigma_1$ : the law
$\ca$ is not commutative.
\end{example}

\subsection{Algebraic structure}

The advantage of the new structures over the ones defined in
section \ref{DP} consists in the omission of the operations over
the groups. Hence, algebraic properties come to light quite naturally.
\\ First, the two laws are associative.

\begin{proposition}\label{as}{\it Associativity}\\
Let $\sigma_1\in \SG_n$, $\sigma_2\in\SG_m$ and $\sigma_3\in\SG_p$
be $3$ permutations
\begin{enumerate}
\item
$\sigma_1\ia(\sigma_2\ia\sigma_3)=(\sigma_1\ia\sigma_2)\ia\sigma_3$
\item $\sigma_1\ca(\sigma_2\ca\sigma_3)=(\sigma_1\ca\sigma_2)\ca\sigma_3$
\end{enumerate}
\end{proposition}
\Proof  1) Set $\eta_1=\sigma_1\ia(\sigma_2\ia\sigma_3)$ and
$\eta_2=(\sigma_1\ia\sigma_2)\ia\sigma_3$. One has
\begin{equation}
\eta_1\circ_Ni=\left\{\begin{array}{ll}
\sigma_1\circ_Ni&\mbox{if }0\leq i\leq n-1\\
\sigma_2\circ_N(i-n)+n&\mbox{if }n\leq i\leq m+n-1\\
\sigma_3\circ_N(i-n-m)+n+m&\mbox{if }n+m\leq i\leq
n+m+p-1\end{array}\right.\nonumber
\end{equation}
for each $0\leq i\leq n+m-1$, and the same holds for $\eta_2\circ_Ni$. It follows that $\eta_1=\eta_2$.\\
2) The strategy is the same. First, we set
$\eta_1=\sigma_1\ca(\sigma_2\ca\sigma_3)$ and
$\eta_2=(\sigma_1\ca\sigma_2)\ca\sigma_3$. The action of $\eta_1$
can be computed as follows
\begin{equation}
\eta_1\circ_N(i+ni')=\sigma_1\circ_Ni+n(\sigma_2\ca\sigma_3)\circ_Ni'=\sigma_1\circ_Ni+n\sigma_2
\circ_Nj+nm\sigma_3\circ_Nk\nonumber
\end{equation}
where $0\leq i\leq n-1$, $0\leq i'\leq mp-1$, $0\leq j\leq m-1$
and $0\leq k\leq p-1$.

On the other hand, the action of $\eta_2$ is
\begin{equation}
\eta_2\circ_N(k'+nmk)=(\sigma_1\ca\sigma_2)\circ_Nk'+nm\sigma_3\circ_Nk=\sigma_1\circ_Ni+n\sigma_2
\circ_Nj+nm\sigma_3\circ_Nk\nonumber
\end{equation}
where $0\leq i\leq n-1$, $0\leq j\leq m-1$, $0\leq k\leq p-1$ and
$0\leq k'\leq nm-1$. Hence, $\eta_1\circ_Ni=\eta_2\circ_Ni$ for
$0\leq i\leq nmp-1$ and $\eta_1=\eta_2$. \cqfd

From example \ref{ex1} and \ref{ex2}, neither
$\rightarrow\hspace{-15pt}+\  $ nor $\ca$ is commutative. But, one
has the property of left distributivity.
\begin{proposition}{\it Semi-distributivity}\\
Let $\sigma_1\in \SG_n$, $\sigma_2\in\SG_m$ and $\sigma_3\in\SG_p$
be three permutations
\[\sigma_1\ca(\sigma_2\ia\sigma_3)=(\sigma_1\ca\sigma_2)\ia(\sigma_1\ca\sigma_3) \]
\end{proposition}
\Proof We use the same method as in the proof of proposition
\ref{as}. First, let us define
$\eta_1=\sigma_1\ca(\sigma_2\ia\sigma_2)$ and
$\eta_2=(\sigma_1\ca\sigma_2)\ia(\sigma_1\ca\sigma_3)$. The action
of $\eta_1$ is
\begin{equation}\label{eta1dis}
\eta_1\circ_N(i+nj)=\eta_1\circ_Ni+n(\sigma_2\ia\sigma_3)\circ_Nj
=\left\{\begin{array}{ll}\sigma_1\circ_Ni+n\sigma_2\circ_Nj&\mbox{if
}0\leq j\leq
m-1\\\sigma_1\circ_Ni+n\sigma_3\circ_N(j-m)+m&\mbox{if }m\leq
j\leq p+m-1\end{array}\right.
\end{equation}
where $0\leq i\leq n-1$ and $0\leq j\leq m+p-1$.\\ On the other
hand, one has \begin{equation}\label{eta2dis}
\eta_2\circ_Nk=\left\{\begin{array}{ll}
(\sigma_1\ca\sigma_2)\circ_Nk&\mbox{if }0\leq k\leq nm-1\\
(\sigma_1\ca \sigma_3)\circ_N(k-nm)+nm&\mbox{if } nm\leq k\leq
n(m+p)-1\end{array}\right. .
\end{equation}
If $0\leq k\leq mn-1$, we set $k=i+nj$ where $0\leq i\leq n-1$ and
$0\leq j\leq m-1$. Hence,
\begin{equation}\label{eta2_1}
(\sigma_1\ca \sigma_2)\circ_Nk=\sigma_1\circ_Ni+n\sigma_2\circ_Nj.
\end{equation}
Similarly, if $nm\leq k\leq n(m+p)-1$, we set $(k-nm)=i+nj$ where
$0\leq i\leq n-1$ and $0\leq j\leq p-1$. Hence,
\begin{equation}\label{eta2_2}
(\sigma_1\ca\sigma_3)\circ_N(k-nm)+nm=\sigma_1\circ_Ni+n(\sigma_3\circ_N(j-m)+m).
\end{equation}
Substituting (\ref{eta2_1}) and (\ref{eta2_2}) in (\ref{eta2dis}),
one recovers the right hand side of (\ref{eta1dis}). It follows
immediately that $\eta_1=\eta_2$. \cqfd

The two laws can be extended by linearity to the graded vector
space $\bigoplus_{n\geq 0}{\mathbb Q}[\SG_n]$ and  endow this
space with a structure of 2-associative algebra. In the next
section, we construct a product $\star$ in $Sym$ (the algebra of
symmetric functions) defined on the power sums and appearing when
one examines the cycle index polynomial of a Cartesian product.
This product is the image of $\ca$ under a particular morphism. We
will prove that this last property implies the associativity and
the distributivity of $\star$ over $\times$ (the natural product
in $Sym$) and $+$.

\section{Cycle index algebra}\label{S3}
\subsection{Cartesian product in $Sym$}
We first construct a 2-associative morphism $\SGS\mapsto Sym$ (a 2-associative algebra is just a vector space
equipped with 2 associative laws \cite{loday1}).\\
The arrow maps a permutation $\sigma\in \SG_n$ to a product of power sums. For $j\geq 1$, let
$c_j(\sigma)$ be the number of cycles in $\sigma$ of length $j$ and set
\begin{equation}
\Zy(\sigma)=\prod_{j=0}^{\infty}\psi_j^{c_j(\sigma)}
\end{equation}
where $\psi_i$ denotes the $i$th power sum symmetric function. We claim
that $\Zy$ is a morphism mapping $\cs$ to $\times$ (the usual product in $Sym$) and that $\ca$ is compatible
with $\Zy$ to the extent that there exists an associative law on $Sym$
such that $\Zy$ is also a morphism mapping it to $\ca$. This second law is
given on the power sums basis by
\begin{equation}\label{defstar}
\prod_{1\leq i\leq \infty} \psi_i^{\alpha_i}\star
\prod_{1\leq j\leq \infty}\psi_j^{\beta_j}=
\prod_{1\leq i,j\leq \infty} \psi_{i\vee
j}^{\alpha_i\beta_j (i\wedge j)}
\end{equation}
(the sequences $(\alpha_i)_{i\geq 1},\ (\beta_j)_{j\geq 1}$ have finite support). It is straightforward to check that

\begin{proposition}\label{sd}
i) The mapping $\Zy: \SGS\mapsto Sym$ is a morphism of 2-associative algebras sending the
two laws $\cs;\ \ca$ respectively to $\times;\ \star$ (recall that $\times$ denotes the usual product of $Sym$).\\ More
precisely, for $\sigma,\ \tau\in \sqcup_{n\geq 0}\SG_n=\SG$ one has
\begin{equation}\label{2assmorph}
\Zy(\sigma\cs \tau)=\Zy(\sigma)\Zy(\tau)\ ;\ \Zy(\sigma\ca \tau)=\Zy(\sigma)\star\Zy(\tau)
\end{equation}
ii) The law $\star$ is associative, commutative and distributive over $\times$.
\end{proposition}

\Proof i) For the first relation of \mref{2assmorph}, one just notices that
$c_j(\sigma\cs \tau)=c_j(\sigma)+c_j(\tau)$. For the second
relation, one observes that the Cartesian product of a $i$-cycle
and a $j$-cycle produces $i\wedge j$ cycles of length $i\vee j$.
Thus, one has
$c_r(\sigma\ca \tau)=\sum_{p\vee q=r}(p\wedge q) c_p(\sigma)c_q(\tau)$, whence \mref{2assmorph}.\\ ii) When
$\sigma\in \SG_n$ is a cycle of maximum length, one has
$\Zy(\sigma)=\psi_n$, hence the image of $\Zy$ contains also all
the products of power sums and we get $Im(\Zy)=Sym$. Then, by
proposition \ref{sd}(i), $\star$ is distributive on the left
over $\times$. Complete distributivity follows from
commutativity of $\star$, which straightforwardly follows from the definition.
\cqfd

The following structural result goes into particulars of the distributivity of $\star$ over $\times$.

\begin{proposition}
Let $P$ be the set of products of power sums (i.e. $P=\{\prod_{i=1}^\infty \psi_i^{\al_i}\}_{(\al_i)_{i\geq 1}\in \N^{\N^*}}$). Then
$P$ is closed by $\times$ and $\star$ and more precisely $(P,\times,\star)$ is isomorphic to a subsemiring of the $\Z$-algebra
$\Z[\N^{\p}]$ of the monoid $(\N^{\p},sup)$ (where $\p$ stands for the set of prime numbers).
\end{proposition}

\Proof The fact that $P$ is closed by $\times$ and $\star$ follows from the definition and \mref{defstar}. Now $P$ contains
the two units ($1$ and $\psi_1$), therefore (as a consequence of the properties established for the laws $\times,\ \star$)
it is a semiring.
For every $p\in \p$ and $n\in \N^*$, let $\nu_p(n)$ be the exponent of $p$ in the decomposition of $n$ in prime factors
($n=\prod_{p\in \p}p^{\nu_p(n)}$). Define an arrow $\phi: P\rightarrow \Z[(\N^{\p}]$ by
\begin{equation}\label{defphi}
\phi(\prod_{1\leq i\leq \infty} \psi_i^{\alpha_i})=\sum_{1\leq i\leq \infty} i\al_i (p\mapsto \nu_p(i)).
\end{equation}
As $\phi(m_1m_2)=\phi(m_1)+ \phi(m_2)$ by construction \mref{defphi}, it suffices to prove that\\
$\phi(\psi_i\star\psi_j)=\phi(\psi_i)\times_s \phi(\psi_j)$ where $\times_s$ stands for the product in
$\Z[(\N^{(\p)},sup)]$. But
\begin{eqnarray*}
\phi(\psi_i\star\psi_j)=\phi(\psi_{i\vee j}^{i\wedge j})=(i\wedge j)\phi(\psi_{i\vee j})=(i\wedge j)(i\vee j)(p\mapsto \nu_p(i\vee j))=\\
 (i\wedge j)(i\vee j)(p\mapsto sup(\nu_p(i),\nu_p(j)))=
ij (p\mapsto sup(\nu_p(i),\nu_p(j)))=
\phi(\psi_i)\times_s \phi(\psi_j).
\end{eqnarray*}
The arrow being clearly into the claim is proved.
\cqfd

\subsection{Cycle index polynomial}
Let $\SG=\bigsqcup_{n\geq 0}\SG_n$ be the disjoint union of all
the symmetric groups and
$\SG_{sg}=\bigcup_{n\geq0}\left(\SG_{n}\right)_{sg}$ be the set
of all the subgroups of all symmetric groups
({\it i.e.} the set of all permutation groups over some interval
$[1..n]$). For simplicity, we identify a permutation group
$G\in\left(\SG_{n}\right)_{sg}$ with its action
$(G,\{0,\dots,n-1\})$ (see section \ref{DP}). Laws $\ia$ and $\ca$
can be  defined over $\SG_{sg}$ by
\begin{equation}
G_1\ia G_2:=(G_1\times G_2,\{0,\dots,n+m-1\})
\end{equation}
where $G_1$ acts on $\{0,\dots,n-1\}$ and $G_2$ acts on
$\{n,\dots,n+m-1\}$ and
\begin{equation}
G_1\ca G_2:=(G_1\times G_2,\{0,\dots,nm-1\})
\end{equation}
where the action on $\{0,\dots,nm-1\}$ is given by
$(\sigma_1,\sigma_2)k=\phi^{-1}((\sigma_1,\sigma_2)\phi(k))$, the
map $\phi$ being the bijection
$\phi:\{0,\dots,nm-1\}\rightarrow\{0,\dots,n-1\}\times\{0,\dots,m-1\}$
defined by $\phi(i+nj)=(i,j)$ if $0\leq i\leq n-1$ and $0\leq
j\leq m-1$ and $(\sigma_1,\sigma_2)(i,j)=(\sigma_1i,\sigma_2j)$.
Note that both $\ia$ and $\ca$ are associative but $\ca$ is not
distributive over $\ia$.\\
Let $Z:\SG_{sg}\rightarrow Sym$ be defined by
\begin{equation}
Z(G)=\Zy\left({1\over|G|}\sum_{\sigma\in G}\sigma\right).
\end{equation}
{\it Poly\`a's cycle index polynomial} of $G$ is defined to be $Z(G)$.
\begin{example} \rm
\begin{enumerate}
\item The cycle index of the symmetric group $\SG_n$ is $Z(\SG_n)=h_n$.
\item The cycle index of the alternating group $A_n$ is
$Z(A_n)=h_n+e_n$.
\end{enumerate}
Here $h_n$ (resp. $e_n$) denotes a complete (resp. elementary)
symmetric function. These examples appear as exercices in
\cite{macdo} (ex.9 p 29).
\end{example}
Since $\Zy$ is a morphism of $2$-associative algebra, one recovers
the classical relations (see \cite{Cam1})
\begin{eqnarray}
Z(G_1\ia G_2)=Z(G_1)Z(G_2)\\ Z(G_1\ca G_2)=Z(G_1)\star Z(G_2)
\end{eqnarray}
\begin{example} \rm
\begin{enumerate}
\item The cycle index polynomial of the Intransitive product of
two symmetric groups $\SG_n$ and $\SG_m$ is
$$Z(\SG_n\ia\SG_m)=h_nh_m.$$
\item The cycle index polynomial of the Cartesian product of two
symmetric groups $\SG_n$ and $\SG_m$ is
$$Z(\SG_n\ca\SG_m)=h_n\star h_m=\sum_{|\lambda|=n,\atop
|\rho|=m}m_\lambda\star m_\rho=\sum_{|\lambda|=n,\atop
|\rho|=m}\frac1{z_\lambda
z_\rho}\prod_{i,j}\psi_{\lambda_i\vee\rho_j}^{\lambda_i\wedge\rho_j},$$
where $m_\lambda$ denotes a monomial symmetric function and
$z_\lambda=\prod i^{n_i}n_i!$ if $n_i$ is the number of parts of
$\lambda$ equal to $i$.
\end{enumerate}
\end{example}

\subsection{Enumeration of a type of Feynman diagrams related to the Quantum Field Theory of partitions}
The cycle index polynomials are classic tools used in combination with Poly\`a's theorem, for the
 enumeration of unlabelled objects. Let us
recall the general process. Consider a permutation group $G$
acting on a finite set $X=\{x_1,\cdots,x_n\}$. Let
$L=\{l_0,\dots,l_p,\dots\}$ (possibly infinite) be another set,
and $f:X\rightarrow L$. The {\it type} $t(f)$ of $f$ is the vector
$(i_0,\dots, i_p,\dots)$ where $i_k$ is the number of elements of
$X$ whose image by $f$ is $l_k$. The {\it shape} $s(f)$ of $f$ is
the partition obtained by sorting in the decreasing order $t(f)$
and erasing the zeroes. For example, a function $f$ having the type
$t(f)=(0,1,0,9,1,2,0,\dots,0,\dots)$ has  the shape
$s(f)=(9,2,1,1)$. The number $d^s_\lambda(G,L)$ of $G$-classes on
$L^X$ with the shape $\lambda$ is the coefficient of $m_\lambda$
in the expansion of $Z(G)$ in the basis of monomial symmetric
functions:
\begin{equation}
Z(G)=\sum_\lambda d^s_\lambda(G,L)m_\lambda .
\end{equation}
Now, let us apply this method to enumerate the Feynman diagrams arising
in the expansion of formula (\ref{QFT0}). These diagrams are
bipartite finite graphs with no isolated vertex, and edges weighted
with integers. First, we enumerate all bipartite finite graphs
with edges weighted with integers. Let $n$ and $m$ be the numbers
of vertices in each of the two parts. We consider the edges as
a function $e$ from $\{0,\dots,n-1\}\times \{0,\dots,m-1\}$ to
$\mathbb N$. The type (resp. the shape) of a graph is the type
(resp. the shape) of its edges, {\em i.e.}  $t(e)$ (resp.
$s(e)$). The number $d_\lambda(n,m)$ of graphs with type $\lambda$
is equal to the number of orbits with type $\lambda$, for the action of $\SG_n\ca\SG_m$ on ${\mathbb
 N}^{\{0,\dots,n-1\}\times \{0,\dots,m-1\}}$.
Hence, the generating function of the shape is
\begin{equation}\label{gnm1}
g(n,m):=\sum_{\lambda}d^s_\lambda(n,m)m_\lambda=Z(\SG_n)\star
Z(\SG_m)
 \end{equation}
Specializing the symmetric functions appearing in (\ref{gnm1}) to
the alphabet $\{y_0,\dots,y_k,\dots,\}$, the coefficient $d_I
^t(n,m)$ of $\prod y_k^{i_k}$ in the expansion of $g(n,m)$ is
equal to the number of graphs with type $I=(i_0,\dots,i_k,\dots)$,
\begin{equation}
g(n,m)=\sum_{I=(i_0,\dots,i_p,\dots)}d^t_I(n,m)\prod_{k=0}^\infty
y_k^{i_k}.
\end{equation}
Note that one can enumerate graphs having edges weighted with
integers less than or equal to $p$ by specializing to the finite
alphabet $\{y_0,\dots,y_p\}$.

Let us define the generating series of the type of our Feynman
diagrams
\begin{equation}
F(n,m):=\sum_{I=(i_0,\dots,i_p,\dots)}f^t_I(n,m)\prod_{k=0}^\infty
y_k^{i_k},
\end{equation}
where $f^t_I(n,m)$ denotes the number of Feynman diagrams of type
$I$. Remark that $F(n,m)$ is a symmetric function over the
alphabet $\{y_1,\dots,y_p,\dots\}$ but not over
$\{y_0,\dots,y_p,\dots\}$.
\begin{example}\label{exF}\rm
Let us give the first examples of generating series for weight in
$\{0,1,2\}$.
\begin{enumerate}
\item $F(1,1)=y_1+y_2$
\item $F(2,1)=F(1,2)=y_1^2+y_1y_2+y_2^2$
\item
$F(2,2)=y_0^2y_1^2+y_0^2y_2^2+y_0^2y_1y_2+y_0y_1^3+3y_0y_1^2y_2+3y_0y_1y_2^2+y_0y_2^3+y_1^4+y_1^3y_2+
3y_1^2y_2^2+y_1y_2^3+y_2^4$
\end{enumerate}
One can remark that under this specialization,
\[F(2,2)+F(2,1)y_0^2+F(1,2)y_0^2+F(1,1)y_0^3+y_0^4=3m_{22}+m_4+3m_{211}+m_{31}=g(2,2).\]
\end{example}
The latter equality could be stated in a more general setting.
\begin{theorem}\label{enum}
One has the following decomposition of the cycle index polynomial.
\begin{equation}
Z(\SG_n\ca\SG_m)=y_0^{nm}+\sum_{(1,1)\leq_{lex} (k,p)\leq_{lex}
(n,m)}F(k,p)y_0^{nm-kp}.
\end{equation}
%
\end{theorem}
\Proof It suffices to remark that a bipartite graph is either a
graph without isolated vertex or the union of some isolated vertex and a smaller bipartite graph. \cqfd
\\
This yields a nice induction formula for the $F(n,m)$'s.
\begin{example}\rm
From theorem \ref{enum}, one has
\[F(3,2)=Z(\SG_3\ca\SG_2)-F(3,1)y_0^3-F(2,2)y_0^2-F(2,1)y_0^4-F(1,2)y_0^4-F(1,1)y^5_0-y_0^6.\]
From example \ref{exF}, it suffices to compute
$F(3,1)=y_1^3+y_2^3$ to enumerate Feynman diagrams whose edges
are weighted by $0$, $1$ or $2$. After simplification, one obtains
\[\begin{array}{rcl}
F(3,2)&=&y_2^6+y_2^5y_1+3y_2^4y_1+3y_2^4y_1y_0+2y_2^4y_0^2+3y_2^3y_1^3+6y_2^3y_1^2y_0+5y_2^3y_1y_0^2\\
&&+y_2^3y_0^3+3y_2^2y_1^4+3y_2^2y_1^3y_0+8y_2^2y_1^2y_0^2+3y_2^2y_1y_0^3+y_2y_1^5+3y_2y_1^4y_0+5y_2y_1^3y_0^2
\\&&+3y_2y_1^2y_0^3+y_1^6+y_1^5y_0+y_1^3y_0^3+2y_1^2y_0^4.
\end{array}\]
For example, there are $8$    $(2,2,2)$- Feynman diagrams:
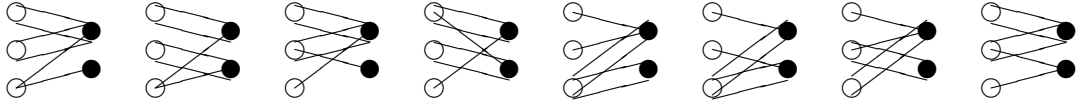
\begin{center}
\setlength{\unitlength}{0.5mm}
\[\begin{array}{cccccccc}
\begin{picture}(30,30)
\put(25,10){\circle*{5}} \put(25,20){\circle*{5}}

\put(5,5){\circle{5}} \put(5,15){\circle{5}}
\put(5,25){\circle{5}}
\put(5,27){\line(4,-1){20}} \put(5,22){\line(4,-1){20}}
\put(5,17){\line(4,1){20}} \put(5,12){\line(4,1){20}}
\put(5,5){\line(4,1){20}} \put(5,5){\line(4,3){20}}

\end{picture}&\begin{picture}(30,30)
\put(25,10){\circle*{5}} \put(25,20){\circle*{5}}

\put(5,5){\circle{5}} \put(5,15){\circle{5}}
\put(5,25){\circle{5}}
\put(5,27){\line(4,-1){20}} \put(5,22){\line(4,-1){20}}
\put(5,17){\line(4,-1){20}} \put(5,12){\line(4,-1){20}}
\put(5,5){\line(4,1){20}} \put(5,5){\line(4,3){20}}
\end{picture}&
\begin{picture}(30,30)
\put(25,10){\circle*{5}} \put(25,20){\circle*{5}}

\put(5,5){\circle{5}} \put(5,15){\circle{5}}
\put(5,25){\circle{5}}
\put(5,27){\line(4,-1){20}} \put(5,22){\line(4,-1){20}}
\put(5,17){\line(4,1){20}} \put(5,12){\line(4,1){20}}
\put(5,5){\line(4,3){20}} \put(5,15){\line(4,-1){20}}

\end{picture}&\begin{picture}(30,30)
\put(25,10){\circle*{5}} \put(25,20){\circle*{5}}

\put(5,5){\circle{5}} \put(5,15){\circle{5}}
\put(5,25){\circle{5}}
\put(5,27){\line(4,-1){20}} \put(5,22){\line(4,-1){20}}
\put(5,17){\line(4,-1){20}} \put(5,12){\line(4,-1){20}}
\put(5,5){\line(4,3){20}} \put(5,25){\line(4,-3){20}}
\end{picture}&
\begin{picture}(30,30)
\put(25,10){\circle*{5}} \put(25,20){\circle*{5}}

\put(5,5){\circle{5}} \put(5,15){\circle{5}}
\put(5,25){\circle{5}}
\put(5,8){\line(4,3){20}} \put(5,3){\line(4,3){20}}
\put(5,7){\line(4,1){20}} \put(5,2){\line(4,1){20}}
\put(5,25){\line(4,-1){20}} \put(5,15){\line(4,1){20}}
\end{picture}&\begin{picture}(30,30)
\put(25,10){\circle*{5}} \put(25,20){\circle*{5}}

\put(5,5){\circle{5}} \put(5,15){\circle{5}}
\put(5,25){\circle{5}}
\put(5,8){\line(4,3){20}} \put(5,3){\line(4,3){20}}
\put(5,7){\line(4,1){20}} \put(5,2){\line(4,1){20}}
\put(5,25){\line(4,-1){20}} \put(5,15){\line(4,-1){20}}
\end{picture}&\begin{picture}(30,30)
\put(25,10){\circle*{5}} \put(25,20){\circle*{5}}

\put(5,5){\circle{5}} \put(5,15){\circle{5}}
\put(5,25){\circle{5}}
\put(5,8){\line(4,3){20}} \put(5,3){\line(4,3){20}}
\put(5,17){\line(4,-1){20}} \put(5,12){\line(4,-1){20}}
\put(5,25){\line(4,-1){20}} \put(5,15){\line(4,1){20}}
\end{picture}&\begin{picture}(30,30)
\put(25,10){\circle*{5}} \put(25,20){\circle*{5}}

\put(5,5){\circle{5}} \put(5,15){\circle{5}}
\put(5,25){\circle{5}}
\put(5,27){\line(4,-1){20}} \put(5,22){\line(4,-1){20}}
\put(5,17){\line(4,1){20}} \put(5,12){\line(4,1){20}}
\put(5,15){\line(4,-1){20}} \put(5,5){\line(4,1){20}}
\end{picture}
\end{array}\]
\end{center}
\end{example}
\section{Non commutative realizations}\label{S4}
\subsection{Free quasi-symmetric cycle index algebra}

Let $(A,<)$ be an ordered alphabet and $w\in A^*$ a word of length $n$. One denotes by $Std(w)$, the permutation $\si\in \SG_n$
defined by
\begin{equation}
\si(i)=(\textrm{Number of letters $= w[i]$ in $w[1..i]$ + number
of letters $< w[i]$ in $w$})
\end{equation}
Recall that the algebra $\FQSym$ is defined by one of its bases, indexed by $\SG$ and defined as follows
\begin{equation}
\F_\sigma=\sum_{Std(w)=\sigma^{-1}}w\in  \ncs{\Z}{A}
\end{equation}
In \cite{nc6}, it is shown that $\FQSym$ is freely generated by
the $\F_\sigma$ where $\sigma$ runs over the connected
permutations (see \cite{Com1}) ({\it i.e.} permutations such that
$\sigma([1,k])\neq[1,k]$ for each $k$). The algebra $\FQSym$ is
spanned by a linear basis, $\{\F^\sigma\}_{\sigma\in \mathfrak S}$, whose product implements the
Intransitive action $\ia$ :
\begin{equation}
\F^\sigma=\F_{\sigma_1}\cdots \F_{\sigma_n}
\end{equation}
where $\sigma=\sigma_1\ia\cdots\ia\sigma_n$ is the maximal
factorisation of $\sigma$ in connected permutations. As a
consequence of this definition, one has
\begin{equation}\label{multF}
\F^\sigma \F^\tau=\F^{\sigma{\rightarrow\hspace{-10pt}+}\ \tau}.
\end{equation}
This naturally induces an isomorphism of algebras
\begin{eqnarray}
\underline\Zy:\left(\bigoplus_{n\geq 0}{\mathbb
Q}[\SG_n],\ia,+\right)&\rightarrow&(\FQSym,.,+)\nonumber\\
\sigma&\mapsto &\F^\sigma.
\end{eqnarray}
One defines the product $\star$ on $\FQSym$ by $\F^\sigma\star
\F^\tau:=\F^{\sigma{\nearrow\hspace{-8pt}\searrow} \tau}$. By this
way, $\underline\Zy$ becomes a morphism of $2$-associative
algebras. Furthermore, $\star$ is associative, distributive over
the sum and semi-distributive over the shifted concatenation.

\subsection{Free quasi-symmetric Poly\`a cycle index polynomial}

 Let $G$ be a permutation group. The {\it free
quasi-symmetric Poly\`a cycle index polynomial} of $G$ is its
image by  $\underline Z:\SG_{sg}\rightarrow \FQSym$ defined by
\begin{equation}\label{nccip}
\underline Z(G):=\underline\Zy\left({1\over|G|}\sum_{\sigma\in
G}\sigma\right)\F^\si.
\end{equation}
\begin{note}
There is another basis of $\FQSym$ indexed by permutations, namely $\{\G^\sigma\}_{\sigma\in \mathfrak S}$. It is obtained
by setting $\G_\si=\F_{\si^{-1}}$ and applying the same construction as
above \mref{multF} to get a basis multiplicative with
respect to $\cs$, then
\begin{equation}
\G^\sigma=\G_{\sigma_1}\cdots \G_{\sigma_n}
\end{equation}
where $\sigma=\sigma_1\ia\cdots\ia\sigma_n$ is the maximal
factorisation of $\sigma$ into connected permutations. In this
case, $\sigma^{-1}$ splits maximally into
$\sigma^{-1}_1\ia\cdots\ia\sigma^{-1}_n$, so one has also
$\G^\sigma=\F^{\sigma^{-1}}$ and formula \mref{nccip} can be rewritten
\begin{equation}\label{nccip}
\underline Z(G):=\underline\Zy\left({1\over|G|}\sum_{\sigma\in
G}\sigma\right)\G^\si.
\end{equation}
\end{note}
The polynomial $\underline Z(G)$ has properties similar to that of
$Z(G)$, in particular regarding the laws $\ia$ and $\ca$.
\begin{proposition}
Let $G_1, G_2\in \SG_{sg}$ be two permutation groups, one has
\begin{enumerate}
\item $\underline Z(G_1\ia G_2)=\underline Z(G_1)\underline
Z(G_2)$.
\item $\underline Z(G_1\ca G_2)=\underline Z(G_1)\star\underline
Z(G_2)$.
\end{enumerate}
\end{proposition}
Consider the morphism, $z:\FQSym\rightarrow Sym$ defined by
$z(\F^\sigma)=\Zy(\sigma)$. Note that it is not a morphism of Hopf
algebra because $z(\F^{231})=\psi_3$.\\ The following diagram is
commutative
\begin{eqnarray}
\SG_{sg}&\displaystyle\mathop{\longrightarrow}^{\underline
Z}&\FQSym\nonumber\\ {\small Z}\downarrow&\displaystyle\mathop{\
}^z\swarrow&\uparrow{\small \underline\Zy}\\
Sym&\displaystyle\mathop{\longleftarrow}_\Zy&\bigoplus_{n\geq 0}{\mathbb
Q}[\SG_n]\nonumber
\end{eqnarray}
\begin{example}\rm
\begin{enumerate}
\item The free quasi-symmetric cycle index of $\SG_n$ is
$${\bf H}_n:=\underline Z(\SG_n)=\frac1{n!}\sum_{\sigma\in
\SG_n}\F^\sigma.$$ One can consider it as a free quasi-symmetric
analogue of the complete symmetric function $h_n$: indeed $z({\bf
H}_n)=Z(\SG_n)=h_n$.
\item One can define free quasi-symmetric analogues of elementary
symmetric functions considering the cycle index polynomial of the
alternative groups: $$ {\bf E}_n:=\underline Z(A_n)-\underline
Z(\SG_n).$$ We get $z({\bf E}_n)=Z(A_n)-Z(\SG_n)=e_n$.
\item The knowledge of analogues of other symmetric functions
should be useful to understand the combinatorics of free
quasi-symmetric cycle index. In particular, it should be
interesting to find free quasi-symmetric  functions whose images
by $z$ are the monomial symmetric functions.
\end{enumerate}
\end{example}

\subsection{Realizations in $\MQSym$}

We will call {\it labelled diagrams} the Feynman diagrams as above but with $p$ white (resp.  $q$ black) spots labelled
bijectively by $[1..p]$ (resp. by $[1..q]$).
When one draws such a diagram, one implicitly assumes that the labelling goes from top to bottom.

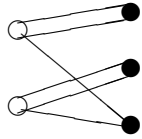
\begin{center}
\setlength{\unitlength}{0.5mm}
\begin{picture}(40,40)
\put(5,10){\circle{5}}
\put(5,30){\circle{5}}

\put(35,5){\circle*{5}}
\put(35,20){\circle*{5}}
\put(35,35){\circle*{5}}

\put(5,32){\line(6,1){30.41}}
\put(5,28){\line(6,1){30.41}}
\put(5,12){\line(3,1){30.41}}
\put(5,08){\line(3,1){30.41}}

\put(6,29){\line(6,-5){30}}
\put(6,9){\line(6,-1){27}}

\end{picture}

Labelled diagram of the matrix  $\begin{pmatrix}2 & 0 & 1\\0 & 2 & 1\end{pmatrix}$.
\end{center}

Now, to such a $p\times q$ {\it labelled diagram} we can associate a
matrix in $\N^{p\times q}$ and this correspondence is one-to-one. The condition that no vertex be isolated is equivalent to
the condition that there be no complete line or column of zeroes, {\em i.e.} the representative matrix is {\it packed} \cite{nc6}.
In the same way, the diagrams are in one-to-one correspondence with the classes of packed matrices under the permutations
of lines and columns as shown below (the vertical arrows are then one-to-one)

\begin{equation}
\begin{CD}
\textrm{Packed matrices}                 @>Class>> \textrm{Classes of packed matrices}\\
@VVV                        @VVV\\
\textrm{Labelled diagrams}                 @>>> \textrm{Diagrams}\\
\end{CD}
\end{equation}
\vspace{5mm}

There is an interesting structure of Hopf algebra (in fact an envelopping algebra) over the diagrams \cite{Marseille} which can
be pulled back in a natural way to labelled diagrams.\\
The correspondence described above allows to construct a new Hopf algebra structure on $\MQSym$ and a Hopf algebra structure on the
space spanned by the classes.

\section{Conclusion}

Other realizations in Hopf algebras seem feasible. For example, let us consider the Hopf algebras of graphs
$GQSym^{110}$ and $GTSym^{110}$ defined in \cite{NT}. An
interesting mapping from $\bigoplus_{n\geq 0}{\mathbb Q}[\SG_N]$ to
$GQSym^{110}$ or $GTSym^{110}$ can be constructed sending each cycle
to an equivalent loop.

\end{document}